# The State of Information and Communication Technology in Hungary – A Comparative Analysis

Peter Sasvari
University of Miskolc, Faculty of Economics, Institute of Business
H3515 Miskolc-Egyetemváros, Hungary
E-mail: iitsasi@uni-miskolc.hu, WebAddress: http://gtk.uni-miskolc.hu/

**Keywords:** information society, ICT, economic sector, electronic presence, electronic marketplace

*A novel comparative research and analysis method is proposed and applied on the Hungarian economic sectors. The question of what factors have an effect on their net income is essential for enterprises. First, the potential indicators related to economic sectors were studied and then compared to the net income of the surveyed enterprises. The data resulting from the comparison showed that the growing penetration of electronic marketpalces contributed to the change of the net income of enterprises in various economic sectors to the extent of 37%. Among all the potential indicators, only the indicator of electronic marketplaces has a direct influence on the net income of enterprises. Two clusters based on the potential indicators were indicated.*

*Povzetek: Predstavljena je primerjalna analiza IKT na Madžarskem.*

## 1 Introduction

The current age is often referred to as the Information Age. This concept was first introduced by Manuel Castells, the best-known theoretician of the information society [1]. The information society is a new, special variant of the existing societies in which producing, processing and distributing information become a fundamental source in the economy.

According to the related literature data, the Information Age began in the second half of the 1950s when, for the first time in history, the number of white-collar workers (engineers, administrative employees etc.) exceeded the number of blue-collar workers ([6]).

One of the main driving forces of the Information Age is the phenomenon called Information and Communication Revolution. Its significance often compared to the agricultural and industrial revolutions taken place in the history of mankind. In important fields of high-end technology (computer technology and telecommunication) not only the robust growth of quality, quantity and performance parameters can be observed but the approximation of these two fields along with the appearance of compound applications can also be detected. These phenomena of the information society cannot only be seen as one of the results of the development of technology but also a coherent system affecting the society as a whole ([3]).

## 2 The characteristics and impacts of information and communication systems

Information and communication technology can be regarded as an universal technological system, which is closely linked to all of the previous systems and creates new, more complex technological systems. ICT's main characterizing function is to assure acquiring, storing, processing, delivering, distributing, handling, controlling, transforming, retrieving and using information. ICT has a different effect on the actors of the economy, including companies, employees and consumers. Nowadays we witness a change of paradigm in the operation of enterprises. They have become a rapidly changing system of independent work groups and projects. Enterprises are characterized by flexible operation and demand for flexible labour force. In this new situation, employees have to leave the traditional patterns and develop a new kind of mentality. If they want to stay afloat in the labour market, they have to be flexible as enterprises are no longer strongly interested in developing the professional knowledge of their employees. Beside the changes experienced in the attitude of enterprises and employees, consumers' behaviour has also been changed essentially by the effect of ICT. As consumers are freed from their isolation by the Internet, they become active and conscious actors in the economy. The relationship between buyers and sellers has changed, it has become harder for sellers to recognize and influence the trends in demand and consumers are better informed than ever before.

Information and communication technology has brought a deep change in the opportunities for consumers compared to the opportunities provided by industrial capitalism. This change was as profound as the one caused by the Industrial Revolution earlier. The new generation of consumers is, first and foremost, well informed, collecting and using other consumers' existing experiences. Companies (especially corporations) previously focused on products and markets, nowadays they concentrate on consumers instead. It is not enough to recognize consumers' problems, identifying the problems in order to solve them is also needed. The opportunities provided by ICT identify actual consumers, based on actual problems occurring during the use of a product. Companies can keep pace with the speed of the development of ICT only by introducing job enrichment. The requirement of versatility can be met only by employees with high-level general education ([3]).

The decrease of the number of strict positions along with the changing requirements of the remaining ones allows employees to acquire new skills but it also stretches their responsibility. Cross-trainings are also organized for the group of employees in order to enable them to perform various tasks. Team-based companies have better problem-solving skills, higher productivity, more efficient use of human resources, more creativity and more innovations when compared to traditional non-team based organizations. Nowadays, when digital information is regarded as the chief mean of production, the efficiency of production is highly dependent upon obtaining and processing information. Based on the achievements of ICT, companies have shaped up the infrastructure of obtaining and processing information, and help their employees to co-operate by compressing time and space. The intention of raising efficiency gave room to virtual teams. By being part of a virtual team, employees do not have to work under the same roof and other employees from outside the company can take part in the work of them.

Nowadays, the majority of changes in work organization, decision mechanisms and corporate organization structures requires enhanced flexibility. Flexibility means quick reaction, the removal of strict limits and the frequently mentioned job enrichment as well as openness for innovations and unconventional answers to the newer and newer challenges. The environmental impulses do not affect the operators of the assembly lines or the workers of call centers through a long chain. Companies were operated centrally from a single headquarters earlier, nowadays managers and workers try to find answers to the current challenges in many local corporate decision nodes. The coordination of numerous independent units is generated by the company as a self-organizing system, and the company's philosophy is determined by the self-organization of independently operating units based on market principles.

## 3   The aim of the research

Based on the considerations presented above, it is not the subject of my paper to answer whether there is a need for ICT or creating the necessary conditions for the information society. The real subject is to measure what economic, social, cultural and environmental effects it has on the society. The rich literature of the information society discusses these aspects in detail. In my work, we take the information society as a normative future plan for Hungary, and we are looking for the answer of what progress has been made in building the information society in the Hungarian economic sectors. We examine the following issues:

- to what extent we can speak about the information society in Hungary nowadays,
- what is the development level of the information society in several economic branches and company sizes compared to each other and to the member states of the European Union,
- how this development level can be measured and calculated,
- how the development level of information and communication technology increases at certain company sizes,
- what trends can be observed in the development process in the individual economic sectors and company sizes.

My examination extends to the static, momentary state of the development level of ICT devices used in the economic branches as well as to their dynamic analysis, expected pace of growth and their qualification. When establishing the aims of the research, there is always the question of how to position the individual parts of the subject. Should they be positioned in a broader subject or should they be selected for further and deeper examination? The former possibility means that we aim to make suggestions by putting the practical analysis into a broader structure. The aim of my research is exactly this, as the information society means a stage representing a new quality, and the changes of the information and communication technology can be observed in every part of our life nowadays.

## 4   The method of the research

Similar problems are raised by the quantification of the various components of the information society as the definition of its concept. There is a wide range of variables that can be measured: a great number of explanatory variables can be listed from the perhaps more easily measurable infrastructural components to the more difficult components related to knowledge and willingness for using information. That is why most analyses use sets of variables and complex indices as there is no easily measurable (one-dimensional) index that would characterize the information society. The examination of the subject is interdisciplinary as it has social and scientific references, so a complex approach was needed when we started processing the literature. We needed to study literature on economics, law, sociology and technology connected to the information society. In consideration of the complexity of the studied subject, we selected several analytical methods and approaches. In the phase of data collection, we relied on the available Hungarian (related reports issued by the Hungarian Central Statistical Office [8]) and international data (Statistical Office of the European Communities [7]) as well, and we managed to process a large amount of secondary information consisting of more than 6.000 items. We extended my research to printed as well as electronic publications and artifacts available on the Internet. The reason for conducting a primary research was to reduce some shortcomings originated from secondary data sources. In fact, it covered an empirical survey among Hungarian companies and enterprises. The questionnaire we used for collecting data on the subject was filled in by 554 respondents altogether, providing nearly 3.000 data records.
As Figure 1 shows, the literature on the development of ICT distinguishes five development stages.

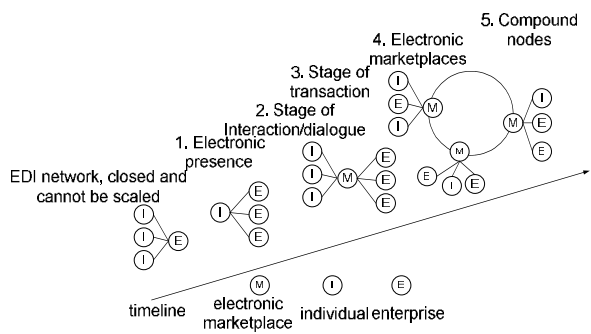

Figure 1: The development stages of information and communication technology [3]

These stages are built upon each other. With the help of the elaborated model, we measured the individual development stages. By averaging the data of the first three development stages, we examined the enterprises' willingness for adaptation. With the help of an own model, which comprises five elements, we analyzed the development and growth of the size categories and economic sectors.
The steps of this procedure are as follows:
- Processing the data of the primary and secondary research,
- Assigning single indicators to individual development stages, calculating potential indicators,
- Calculating the values of potential indicators from single indicators,
- Studying potential indicators,
- Determining potential indicators at the individual development stages.

Then, with the help of the resulting indicators, we performed a cluster analysis, a compound regression analysis, and finally a discriminant analysis on the surveyed economic sectors.

## 5 The results of the analysis of information and communication technology

Clustering is the assignment of a set of observations into subsets so that observations in the same cluster are similar in some sense. The clustering process is successful when the subsets are similar to each-other and different from the elements of other subsets at the same time. Based on theoretical considerations, we decided to make groups of economic activity categories from the five previously defined potential indicators.

As a summary of the results of the cluster analysis, it can be stated that the sectors "Electricity, gas and water supply", "Transport, storage and communication", "Mining and quarrying", "Manufacturing" and "Financial intermediation" belong to the second cluster by better average values. The results of this analysis are presented in Figure 2.

| 1st cluster | 2nd cluster |
|---|---|
| (A) Agriculture, hunting and forestry<br>(F) Construction<br>(G) Wholesale and retail trade; repair work<br>(H) Hotels and restaurants<br>(K) Real estate, renting and business activities<br>(M) Education<br>(N) Health and social work | (C) Mining and quarrying<br>(D) Manufacturing<br>(E) Electricity, gas and water supply<br>(I) Transport, storage and communication<br>(J) Financial intermediation |
| Underdeveloped | Developed |
| relative ||

Figure 2: Two-cluster model of the national economic sectors

We used the path model to study how the potential indicators influence one another and what direct or indirect effect they have on the average net income of the individual economic sectors.

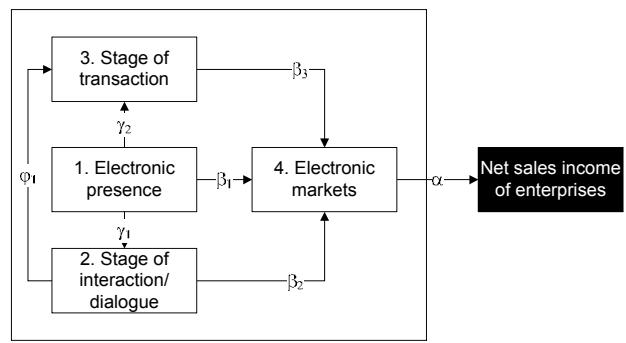

Figure 3: The scheme of the path model of the potential indicators

As it is indicated in Figure 3, the variables presented in the path model are linked with arrows to one another, showing the direction of their relationships. We assumed in my causal model that the potential indicator of electronic presence is the exogenous variable. Based on the arrows starting from it, the potential indicator of electronic presence has an effect on the other potential indicators, also having an indirect effect on the average net income of enterprises in several economic sectors. These paths are called indirect paths by the literature and in my model they show how the effect of the potential indicator of electronic presence takes place through the potential indicators of interaction/dialoge, transaction and electronic markets. The potential indicators of interaction/dialogue and transaction became endogenous variables. Endogenous variables are variables with causal links leading to them from other variables in the model. In other words, endogenous variables have explicit causes within the model. The dependent variable in my model is the average net income of enterprises in economic sectors, the arrows starting from the other variables point at this one but it has no arrow or link pointing back at the other variables.

The aim of setting up a path model was to divide the zero linear correlation between the independent and the dependent variables into two parts. The first part is the effect that the independent variable directly has on the dependent variable, while the second part shows the effect being had on the dependent variable caused by the independent variable through another endogenous variables.

Only the potential indicator of electronic markets has a direct effect on the average net income of enterprises as it is illustrated in Figure 4. However, the effect of the potential indicator of electronic presence is significant as it influences the potential indicator of electronic markets to a great extent. The value of the indirect effect of electronic presence was (87.4%*60.5%) 56.2%. In the table below, a new arrow also appears with a value of 70%, showing the effect of non-specified variables from outside the model on the average net income of enterprises.

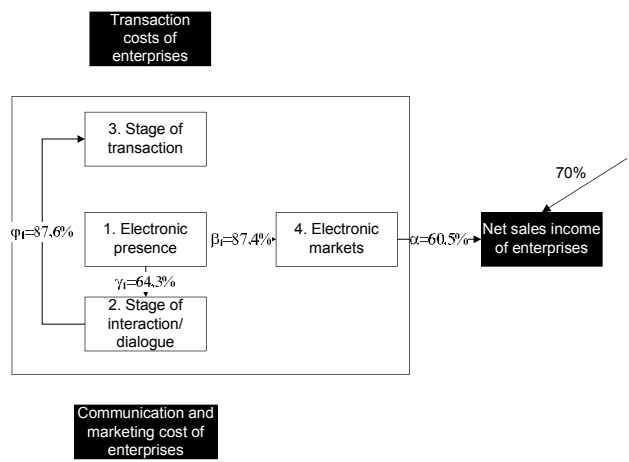

Figure 4: The final path model of the potential indicators

Electronic presence has no direct effect on the potential indicator of transaction. The value of the strength of its indirect effect was (64.3%*87.6%) 56.3%, according to my computation. The model verified the hypothesis according to which electronic presence largely determines interaction/dialogue, it has an indirect effect on transaction and it has the strongest correlation with electronic markets. Before creating the model, we assumed a direct correlation between transaction and electronic markets but we could not verify the existence of the relationship between them. However, the new result of my research was that there was a direct correlation between electronic markets and the average net income of enterprises.

In the early phase of my research, we encountered the problem that there were no explanatory variables in the typology created by cluster analysis. Typologies, different clusters are of a low measurement level so the explanation of their development status is impossible with the formerly used techniques. Discriminant analysis is a useful method to explain a low measurement level variable with another variable of high measurement level. Discriminant analysis is a technique where dependent variables are not metric and are classified between two or more categories whereas independent variables (predictors) are measured on a metric scale. The summary of the methods used together with discriminanat analysis is shown in Figure 5.

|  | Independent variable | |
|---|---|---|
|  | Non-metric | Metric |
| Dependent variable — Non-metric | Crosstabs analysis | Discriminant analysis |
| Dependent variable — Metric | Variant analysis | Correlation, regression analysis |

Figure 5: Partial summary of the methods used for structure analysis, along with discriminant analysis [4]

After completing the cluster analysis, we found that the surveyed economic sectors could be classified into two, then four ICT development levels or clusters. The resulting four-cluster model is illustrated in Figure 6.

| 1.2 cluster | 2.2 cluster |
|---|---|
| (G) Wholesale and retail trade; repair work<br>(K) Real estate, renting and business activities<br>(M) Education | (D) Manufacturing<br>(E) Electricity, gas and water supply<br>(I) Transport, storage and communication<br>(J) Financial intermediation |
| 1.1 cluster | 2.1 cluster |
| (A) Agriculture, hunting and forestry<br>(F) Construction<br>(H) Hotels and restaurants<br>(N) Health and social work | (C) Mining and quarrying |
| Underdeveloped | Developed |
| relative | |

Figure 6: Four-cluster model of the national economic sectors

The following four economic sectors got into the 1.1 cluster: 'Agriculture, hunting and forestry', 'Construction', 'Hotels and restaurants' and 'Health and social work'. The average of the potential indicators to electronic presence, interaction/dialogue, transaction and electronic markets was the lowest in the four clusters.

Four economic sectors were classified into the 1.2 cluster as well: 'Wholesale and retail trade; repair work', 'Real estate, renting and business activities', 'Education' and 'Other community, social and personal service activities'. Examining the data of this cluster, it could be observed that its average values were higher than those of the 1.1 cluster but were lower than the average values of the other two clusters.

Only the 'Mining and quarrying' sector was classified into the 2.1 cluster. In terms of electronic presence and electronic markets, this sector was the most developed compared to the other sectors. This cluster produced the second highest ICT values based on the values of the other potential indicators.

'Manufacturing', 'Electricity, gas and water supply', 'Transport, storage and communication' and 'Financial intermediation' could be found in the 2.2 cluster. The values of interaction/dialogue and transaction were the highest in this cluster comparing to the other ones.

My aim was to get to know the human resource demand of enterprises (the number of the employees regularly using computers), the cost of ICT services or availability (cost of computer-related services) and the amount spent on professional training (the total expenditure on professional training). These three explanatory variables jointly indicate the different ICT development stages, in this case discriminant analysis predicts whether an enterprise belongs to a specific development stage or not. Based on the primary research, it can be stated that education expenses have a more significant effect on belonging to various clusters. As the aim of the discriminant analysis is the classification of cases into groups, the classification table is one of the most important results of the analysis. The table below consists of two parts: the first presents the scores before the grouping took place. The chance of being classified into a cluster is 25% in each group and each cluster weight was different.

| Cluster | Prior | Cases Used in Analysis | |
|---|---|---|---|
| | | Unweighted | Weighted |
| 1.1 | ,250 | 36 | 36,000 |
| 1.2 | ,250 | 4 | 4,000 |
| 2.1 | ,250 | 58 | 58,000 |
| 2.2 | ,250 | 82 | 82,000 |
| Total | 1,000 | 180 | 180,000 |

| | | Cluster | Predicted Group Membership | | | | |
|---|---|---|---|---|---|---|---|
| | | | 1.1 | 2.1 | 2.2 | 1.2 | Total |
| Original | Count | 1.1 | 19 | 0 | 0 | 17 | 36 |
| | | 2.1 | 2 | 1 | 0 | 1 | 4 |
| | | 2.2 | 24 | 0 | 4 | 30 | 58 |
| | | 1.2 | 30 | 1 | 0 | 51 | 82 |
| | % | 1.1 | 52,8 | ,0 | ,0 | 47,2 | 00,0 |
| | | 2.1 | 50,0 | 25,0 | ,0 | 25,0 | 00,0 |
| | | 2.2 | 41,4 | ,0 | 6,9 | 51,7 | 00,0 |
| | | 1.2 | 36,6 | 1,2 | ,0 | 62,2 | 00,0 |
| Cross-validated a | Count | 1.1 | 16 | 0 | 1 | 19 | 36 |
| | | 2.1 | 2 | 0 | 0 | 2 | 4 |
| | | 2.2 | 24 | 1 | 3 | 30 | 58 |
| | | 1.2 | 32 | 1 | 1 | 48 | 82 |
| | % | 1.1 | 44,4 | ,0 | 2,8 | 52,8 | 00,0 |
| | | 2.1 | 50,0 | ,0 | ,0 | 50,0 | 00,0 |
| | | 2.2 | 41,4 | 1,7 | 5,2 | 51,7 | 00,0 |
| | | 1.2 | 39,0 | 1,2 | 1,2 | 58,5 | 00,0 |

Table 1: Classification Results

The actual hit ratio can be seen in the second part, it is given in percentage, its value ranges from 0 to 100. Instead of the lowest possible value, it needs to be compared to the expected hit ratio. The expected hit ration means the hit ratio resulting from random categorization, its value is 25% in the case of four groups.

The classification table is suitable for the evaluation of the results of the discriminant analysis as it shows the ratio of the adequately categorized group membership. The rows make up the categories of the dependent variables and their initially observed values, while the columns of the table constitute the values predicted by the independent variables. The table can be divided into two parts: the upper part of it shows the inital analysis, while its lower part presents the cross validation values. The data are presented in the same way in both parts of the table, they are expressed either in absolute value or in percentage. Analyzing the absolute values of the table, it can be observed that only 19 cases got into the 1.1 cluster from its original 36 cases, while 17 of them got into the 1.2 cluster. Expressing this data in percentage it means that the rate of the adequately categorized cases is 52.8% in the 1.1 cluster, 25% in the 2.1, 6.9% in the 2.2 and 62.2% in the 1.2 cluster. Consequently, the procedure was successful only in the cases of the 1.1 and the 1.2 clusters. SPSS identifies values as adequate hit ratio on the diagonal: if the prediction equals the value of the initial sets of observations then the prediction is perfect and every value is situated on the diagonal. Enterprises were adequately categorized in 41.7% of cases and 37.2% of predictions based on the given variables.

In summary, it can be stated that the first and the fourth clusters are significantly different from the other two clusters, as their hit ratio is above 50% in the case of three independent variables. Examining the results, it can also be observed that these two clusters can hardly be divided in the case of three independent variables.

# 6 Conclusions and suggestions for the practical use of research findings

The most important step of the cluster analysis is to determine the number of clusters. The data show that it is expedient to form two clusters based on the potential indicators. The first cluster comprises eight, while the second comprises five economic activities. As a consequence, those economic sectors belong to the first cluster that use ICT devices less frequently than the national average, while the second cluster contains those economic sectors that can be seen as developed ICT-users.

The multiple regression analysis is the series of regression models built upon each other. Using the regression model, we studied the direct and indirect effect of the potential indicators on each other and the companies' net income in several economic sectors. The only potential indicator affecting a company's net income is the indicator of electronic marketplaces. However, the effect of the electronic presence is significant, since it has a great influence on the potential indicator of electronic marketplaces. During my primary research, we found out that the effect of the non-specified variables out of the regression model on a company's net income is 70%.

The typology carried out by cluster analysis does not contain independent variables. The discrimination analysis helps to explain the values of dependent variables with the help of independent variables. With the clusters showing the given development stages, my aim was to get a better idea on the companies' needs of human resources and on how much is spent on training and ICT services by the given company. Exclusively training expenses have a more significant effect on which cluster a company belong to. It was possible to classify the companies into clusters based on the three independent variables in 42% of the cases.

We could not find a reassuring mathematical and statistical method for studying the effect of the information communication technology on businesses in the literature, that is why we proposed a new research and analysis method that we also used to study the Hungarian economic sectors.

The primary possibility of utilizing the proposed method appears in situation report. We managed to measure the relative (economic sectors correlated to each other) and the absolute (economic sectors correlated to the same ones in a different country) development level of the information communication technology with the help of creating development stages, quality categories and the willingness for adoption belonging to the given development stages.

The secondary possibility for utilization lies in following patterns. The development of ICT is different in several countries, regions and economic sectors. The European Union proposed a strategic framework for its member countries. The main aims of establishing a strategic framework are:
- a single European information space;
- boosting investment and innovation in ICT researches;
- establishing a receptive European information society.

The economy of the United States is regarded as a model economy where two-third of the employees were dealing with information process during working hours in 2000. One of the causes of the massive economic performance in the United States is the highly-developed information processing. If we manage to measure this level of development, a strategy can be formulated in the European Union and in the individual member states in order to catch up with the most developed countries.

The object of the study is generally the national economy of a given country. With the help of the method we have worked out, it is possible to analyze and assess the sections, subsections, divisions, groups and classes of a given national economy. Beside the economic sectors, company sizes and organization forms can also be studied.

## Acknowledgement

We owe my deepest gratitude to my scientific leader, associate professor Pelczné Dr Ildikó Gáll who has been guiding me with great expertise and patience for years, following my professional work with attention and giving me assistance as well as useful advices and suggestions.